\documentclass[%
 reprint,
superscriptaddress,
preprintnumbers,
nofootinbib,
 amsmath,amssymb,
 aps,
 prd,
floatfix,
]{revtex4-2}
\usepackage{mathtools}
\usepackage{graphicx}% Include figure files
\usepackage{dcolumn}% Align table columns on decimal point
\usepackage{bm}% bold math
\usepackage[export]{adjustbox}
\usepackage{aas_macros}
\usepackage{hyperref}% add hypertext capabilities
\hypersetup{colorlinks = true, citecolor = blue, linkcolor = blue}
\DeclarePairedDelimiter{\ceil}{\lceil}{\rceil}
\newcommand{\n}[1]{\mathbf{#1}}

\newcommand{\target}{PSR J0437$-$4715}
\newcommand{\Ti}{$T_{\rm drift}^{({\rm i})}$}
\newcommand{\Tii}{$T_{\rm drift}^{({\rm ii})}$}
\newcommand{\maxL}{${\rm max}(\mathcal{L})$}
\DeclareUnicodeCharacter{2212}{-}
\begin{document}

%\preprint{APS/123-QED}

\title{Search for continuous gravitational waves from PSR J0437$-$4715 with a hidden Markov model in O3 LIGO data}

\author{Andres F. Vargas}
\email{afvargas@student.unimelb.edu.au}
\author{Andrew Melatos}%
\affiliation{School of Physics, University of Melbourne, Parkville, Vic, 3010, Australia}%
\affiliation{OzGrav-Melbourne, Australian Research Council Centre of Excellence
for Gravitational Wave Discovery, Parkville, Victoria, 3010, Australia}

\date{\today}

\begin{abstract}
Results are presented for a semi-coherent search for continuous gravitational waves from the millisecond pulsar \target, using a hidden Markov model to track spin wandering, in LIGO data from the third LIGO-Virgo observing run. This is the first search for~\target~to cover a wide frequency range from $60$ Hz to $500$ Hz and simultanously accommodate random spin deviations from the secular radio ephemeris. Two searches are performed with plausible coherence times of $10$ days and $30$ days, as the frequency wandering time-scale of the gravitational-wave-emitting quadrupole is unknown. The former analysis yields no surviving candidates, while the latter yields five candidates after the veto procedure. The detection statistic of each of the five survivors is mapped as a function of sky position, in preparation for follow-up analyses in the future, e.g. during LIGO-Virgo-KAGRA fourth observing run.
\end{abstract}

\maketitle

\section{Introduction}

Rapidly rotating neutron stars (NSs) are potential sources of continuous gravitational waves (GWs) detectable by terrestrial detectors such as the Advanced Laser Interferometer Gravitational-wave Observatory (LIGO)~\cite{Riles2013,HarryLIGOScientificCollaboration2010,LIGOScientificCollaborationAasi2015,AnderssonFerrari2011}, Advanced Virgo~\cite{AcerneseAgathos2015}, and the Kamioka Gravitational Wave Detector (KAGRA)~\cite{AkutsuAndo2021}. Several mechanisms, such as elastic stresses~\cite{UshomirskyCutler2000,Johnson-McDanielOwen2013}, magnetic gradients \cite{Cutler2002,MastranoMelatos2011,LaskyMelatos2013}, $r$-modes \cite{Heyl2002,ArrasFlanagan2003,BondarescuTeukolsky2009}, or non-axisymmetric circulation of the superfluid interior \cite{PeraltaMelatos2006,vanEysdenMelatos2008,BennettvanEysden2010,MelatosDouglass2015}, can produce an oscillating quadrupole moment which emits continuous GWs at specific multiples of the NS spin frequency $f_{\star}$ \cite{Riles2013}. 

Millisecond pulsars (MSPs) with an accurately measured radio ephemeris, such as \target~\cite{JohnstonLorimer1993, SosaFiscelladelPalacio2021}, are key targets for continuous GWs searches. Non-axisymmetries, supported ellastically or magnetically, can be generated in the NS crust through accretion \cite{HaenselZdunik1990, UshomirskyCutler2000, MelatosPayne2005} and subside with age \cite{WetteVigelius2010}. In addition, the $r$-mode instability may play a role in the spin history of these objects \cite{AnderssonKokkotas1999a,AnderssonKokkotas1999b, Bildsten1998}, with periods of activity ranging from short \cite{levin1999} to long \cite{NayyarOwen2006} timescales. These different avenues for MSPs to maintain an oscillating quadrupole moment have served historically to motivate coherent searches for continuous GWs, using a bar detector \cite{MohantyHeng1998} or terrestrial interferometers \cite{AbbottAbbott2019b,AbbottAbbott2020,TheLIGOScientificCollaborationtheVirgoCollaboration2021a}.

In this paper we search for \target~with an hidden Markov model (HMM) algorithm, which tracks spin wandering \cite{SuvorovaSun2016, AbbottAbbott2017,SuvorovaClearwater2017,AbbottAbbott2019a,TheLIGOScientificCollaborationtheVirgoCollaboration2022}. There are strong astrophysical motivations to target this object. (1) It is nearby, at a distance $D=156.3\pm1.3{\rm~pc}$ measured from parallax to $1\%$ precision \cite{DellerVerbiest2008}, and the GW strain scales $\propto D^{-1}$. (2) It spins rapidly, with $f_{\star} = 173.69 ~{\rm Hz}$ \cite{JohnstonLorimer1993}, and the GW strain scales $\propto f_{\star}^2$. (3) Radiation at $f_{\star}$ and $2f_{\star}$ from a static mountain and $\approx 4f_{\star}/3$ \cite{CarideInta2019} from an $r$-mode falls in a sensitive band of the LIGO detectors, separated from instrumental noise lines \cite{CovasEffler2018}. (4) The orbital elements are known to high accuracy from radio pulsar timing, e.g. the orbital period $P=5.741~{\rm days}$ has an uncertainty of $\sigma_{P}=3\times10^{-7}~{\rm days}$ (one standard deviation) \cite{PereraDeCesar2019}. 

Previous searches for \target, like \cite{AbbottAbbott2019b, AbbottAbbott2020, TheLIGOScientificCollaborationtheVirgoCollaboration2021a}, are confined to narrow ranges of $\sim1~{\rm Hz}$ around $f_{\star}$, and $2f_{\star}$. It is eminently possible that the GW-emitting quadrupole is not locked to the crust, so that the GW signal frequency is displaced significantly from simple rational multiples of $f_{\star}$ and possibly also wanders stochastically with time \cite{MukherjeeMessenger2018}. The HMM algorithm is adept at tracking a displaced and wandering tone. Hitherto, HMM narrow-band searches run in sub-bands of $\sim1~{\rm Hz}$ around multiples of the spin frequency like $f_{\star}$ and $2f_{\star}$ (mountains) and $4f_{\star}/3$ ($r$-modes) \cite{MiddletonClearwater2020,BeniwalClearwater2021,TheLIGOScientificCollaborationtheVirgoCollaboration2021}. However, the HMM runs so fast $(\sim1~{\rm day})$, when the orbital elements are known accurately as for \target, that we elect to cover all options in this paper and search LIGO's most sensitive frequency band from $60~{\rm Hz}$ to $500~{\rm Hz}$, to ensure that nothing is missed. This strategy acknowledges recent theoretical work that $r$-modes may emit far from $4f_{\star}/3$ \cite{IdrisyOwen2015,CarideInta2019}.

The paper is organized as follows. Section~\ref{sectionII:HMM_Algorithm} introduces the HMM scheme, and  the frequency domain matched filter, i.e. the $\mathcal{J}$-statistic, used for the search.  Section~\ref{sectionIII:the_target} briefly describes the target and its parameter space. In Section~\ref{sectionIV:Search_implementation} the search pipeline and threshold procedure are reviewed. Section~\ref{SectionV:Analysis_Results} presents the search results. Astrophysical implications are canvassed briefly in Section~\ref{SectionVI:Conclusions}.

\section{HMM Algorithm}
\label{sectionII:HMM_Algorithm}

The HMM analysis pipeline involves two components. The first, namely the HMM itself, tracks the frequency evolution of the signal from one time-step to the next. It is described in Section~\ref{subsecII:HMM_formalism}. The second, namely the $\mathcal{J}$-statistic, computes the likelihood that a signal (if any) is present in the data between two time-steps as a function of the frequency. It is described in Section~\ref{subsecII:J-statistic}.

\subsection{Formalism}
\label{subsecII:HMM_formalism}

A HMM is characterized by a hidden state variable, $q(t)$, and an observable state variable, $o(t)$. As in previous works \cite{SuvorovaSun2016, AbbottAbbott2017,SuvorovaClearwater2017,AbbottAbbott2019a,TheLIGOScientificCollaborationtheVirgoCollaboration2022} the hidden state variable is the GW frequency $f(t)$. One can have $f(t) \neq f_{\star}(t)$ because (i) different emission mechanisms produce GWs at different multiples of $f_{\star}(t)$ \cite{Riles2013}; (ii) the GW-emitting quadrupole may not corotate exactly with the stellar surface and magnetosphere, whence the electromagnetic emission originates; and (iii) $f(t)$ and $f_{\star}(t)$ may wander stochastically in response to fluctuating internal torques \cite{MukherjeeMessenger2018}, such that the lag $f(t)-f_{\star} \neq 0$ fluctuates as well .

The HMM jumps between discrete hidden state values $\{q_{1},...,q_{N_{Q}}\}$ at discrete epochs $\{t_{0},...,t_{N_{T}}\}$. The probability to jump from $q_{i}$ at $t_{n}$ to $q_{j}$ at $t_{n+1}$ is given by the transition matrix $A_{q_{j}q_{i}}$,  which takes the form

\begin{equation}
    A_{q_{j}q_{i}} = \frac{1}{3} \left( \delta_{q_{j}q_{i-1}}+\delta_{q_{j}q_{i}}+\delta_{q_{j}q_{i+1}} \right),
    \label{Eq_SectionII:transition_matrix_Aqq}
\end{equation}

\noindent where $\delta_{ij}$ is the Kronecker delta. Eq.~(\ref{Eq_SectionII:transition_matrix_Aqq}) describes a piece-wise constant signal model, where $f(t)$ jumps by $0$, or $\pm 1$ frequency bins with equal probability at each epoch. Other signal models are viable too, of course, but tests on synthetic data verify that the performance of the HMM depends weakly on the exact form of $A_{q_{j}q_{i}}$ \cite{QuinnHannan2001,SuvorovaSun2016}.

The Fourier transform of the time-domain data collected by the detectors is used as the observable state variable. The complete data stream, with duration $T_{\rm obs}$, is divided into $N_{T}$ segments of duration $T_{\rm drift}$, following $N_{T}={\rm floor}(T_{\rm obs}/T_{\rm drift})$. The coherence time $T_{\rm drift}$ is chosen judiciously to avoid $f(t)$ wandering more than one frequency bin per time-step. In general HMM searches \cite{AbbottAbbott2017,SuvorovaClearwater2017,AbbottAbbott2019a,TheLIGOScientificCollaborationtheVirgoCollaboration2022} use the observed X-ray flux variability of systems like Scorpius X-1 to infer the stochastic variation in $f(t)$ \cite{MessengerBulten2015,MukherjeeMessenger2018} and make an informed choice of $T_{\rm drift}$. Section~\ref{sectionIV:Search_implementation} expands on the choice of $T_{\rm drift}$ for this search. The likelihood of observing $o_{j}$ if the hidden state variable is $q_{i}$ is encoded in the emission probability matrix $L_{o_{j}q_{i}}$. We use the $\mathcal{J}$-statistic, a frequency domain estimator reviewed in Section~\ref{subsecII:HMM_formalism}, to calculate $L_{o_{j}q_{i}}$~\cite{SuvorovaClearwater2017}. 

The probability the hidden state variable follows a path $Q=\{q(t_{1}),...,q(t_{N_{T}})\}$, given the observations $O=\{o(t_{0}),...,o(t_{N_{T}}) \}$, is the product of the transition and emission probabilities per time-step, namely

\begin{align}
    \Pr(Q\vert O) = \;\;\;&L_{o(t_{N_{T}})q(t_{N_{T}})}A_{q(t_{N_{T}})q(t_{N_{T}-1})}\times\,...\, \nonumber \\ 
    &\times L_{o(t_{1})q(t_{1})}A_{q(t_{1})q(t_{0})}\Pi_{q(t_{0})}, 
    \label{Eq_SectionII:Prob_Q_given_O}
\end{align}

\noindent where $\Pi_{q(t_{0})}$ is the prior probability of starting at $q(t_{0})$ and is taken to be uniform across all initial states, with $\Pi_{q(t_{0})}=1/N_{Q}$.

The Viterbi algorithm searches efficiently for the path $Q^{*}$ that maximizes Eq.~(\ref{Eq_SectionII:Prob_Q_given_O}) \cite{AViterbi1967}. A thorough explanation of the algorithm, in the context of GW searches, is included in Appendix A of Ref.~\cite{AbbottAbbott2017}. As a detection statistic we use the log likelihood of the most likely path given the data, i.e. $\mathcal{L}= \ln\,\Pr(Q^{*}\vert O)$.

\subsection{$\mathcal{J}$-statistic}
\label{subsecII:J-statistic}

The emission probability $L_{o(t_{n})q(t_{n})}$ relates the observed data between two consecutive epochs, $\-\{o(t') \vert t_{n} \leq t' \leq t_{n}+T_{\rm drift}\}$, to the hidden states occupied at the first epoch, $q(t_{n})$. $L_{o(t_{n})q(t_{n})}$ is expressed in terms of the $\mathcal{J}$-statistic \cite{AbbottAbbott2019a,SuvorovaClearwater2017}. The $\mathcal{J}$-statistic extends the traditional $\mathcal{F}$-statistic \cite{JaranowskiKrolak1998}, a matched filter for a biaxial rotor \cite{PrixKrishnan2009}, by tracking the movement of the NS in its binary system and accommodating the orbital Doppler shift of the GW carrier frequency. To track the NS, the $\mathcal{J}$-statistic assumes a circular Keplerian orbit (\target~has orbital eccentricity $e=1.9180\times10^{-5}$ \cite{VerbiestBailes2008}). It is a function of three binary orbital parameters: the orbital period $P$, the orbital phase at a reference time $\phi_{a}$, and the projected semi-major axis $a_{0}$.

The $\mathcal{J}$-statistic is efficient computationally. It executes on graphical processing units (GPU) and ingests prefabricated $\mathcal{F}$-statistic \textit{atoms} \cite{Fatoms} which are independent of the binary orbital parameters. The complete definition of the $\mathcal{J}$-statistic can be found in Section III in Ref.~\cite{SuvorovaClearwater2017}.

\section{PSR J0437$-$4715}
\label{sectionIII:the_target}

\target, discovered in 1993 \cite{JohnstonLorimer1993}, is the brightest nearby millisecond pulsar known. It exists in a binary with a cool hydrogen atmosphere white dwarf~\cite{DurantKargaltsev2012} and travels at $104.71\pm0.95~{\rm km\,s}^{-1}$ \cite{DellerVerbiest2008} through the interstellar medium. As well as the radio ephemeris, the study of this system spans the optical \cite{BellBailes1993}, X-ray \cite{ZavlinPavlov2002}, and ultraviolet \cite{KargaltsevPavlov2004} bands, which translates into accurate electromagnetic values for the search parameters used by the HMM pipeline. Section~\ref{subsecIII:target_params} quotes the sky position and orbital parameters used throughout this paper, while Section~\ref{subsecIII:spin_freq} justifies the frequency, and its derivatives, probed by the analysis. 

\subsection{Sky position and orbital parameters}
\label{subsecIII:target_params}

The $\mathcal{J}$-statistic depends on the right ascension $\alpha$ and declination $\delta$ of the source (through the $\mathcal{F}$-statistic atoms) in addition to the three orbital parameters $P,\,\phi_{a}$, and $a_{0}$. In practice, we rewrite $\phi_{a}$ in terms of the time of passage through the ascending node, $T_{\rm asc}$, via $\phi_{a}=2\pi T_{\rm asc}/P \; ({\rm mod}\,2\pi)$. These parameters have been measured electromagnetically at a reference epoch $T_{0}=55000\,{\rm MJD}$ (June $18$ $2009$)~\cite{PereraDeCesar2019}. Their values and uncertainties are presented in Table~\ref{tab:Target_Params}. 

The electromagnetically measured value $T_{\rm asc, ref}$, dated at $T_{0}$, is propagated forward to the start of O3, $T_{\rm O3,0}=1238166483$ GPS time, following

\begin{equation}
    T_{\rm asc} = T_{\rm asc, ref}+N_{\rm orb}P,
    \label{Eq_SectionIII:propagated_Tascref_to_Tasc}
\end{equation}

\noindent where $P$ is the central value of the orbital period in Table~\ref{tab:Target_Params}, and $N_{\rm orb}={\rm ceil}\left[(T_{\rm O3,0}-T_{0})/P\right]$ is the number of full orbits from $T_{0}$ to $T_{\rm O3,0}$. The propagated uncertainty for $T_{\rm asc}$ is calculated in terms of the uncertainty in the period, $\sigma_{P}$, as

\begin{equation}
    \sigma_{T_{\rm asc}}=\sqrt{\sigma_{T_{\rm asc},0}^{2}+(N_{\rm orb}\sigma_{P})^{2}},
    \label{Eq_SectionIII:uncertainty_compund_Tasc}
\end{equation}

\noindent where $\sigma_{T_{\rm asc},0}$ is the uncertainty of $T_{\rm asc, ref}$. Eq.~(\ref{Eq_SectionIII:uncertainty_compund_Tasc}) yields $\sigma_{T_{\rm asc}}=55\,{\rm s}$. This is the value included in Table~\ref{tab:Target_Params}. 

Apart from $T_{\rm asc}$, the remaining orbital parameters are known to high precision, with the uncertainty in $P$ satisfying $\sigma_{P} \leq 10^{-1}$s, and the uncertainty in $a_{0}$ satisfying $\sigma_{a_{0}} \leq 10^{-7}$ lt-s. These values and their uncertainties are obtained from the second data release of the International Pulsar Timing Array~\cite{PereraDeCesar2019}.

\subsection{Spin frequency}
\label{subsecIII:spin_freq}

The spin frequency is measured to high accuracy by radio timing, with $f_{\star}=173.6879457375201\pm9\times10^{-13}$Hz at $T_{0}$. In this paper, we elect to search a much wider range of $60$ Hz to $500$ Hz, where LIGO is most sensitive, to account for $f(t) \neq f_{\star}(t)$ as discussed in Section~\ref{subsecII:HMM_formalism}. The frequency mismatch is likely to satisfy $\vert f(t)- \eta f_{\star}(t) \vert \ll f_{\star}(t)$, where $\eta$ is a simple rational number (e.g. $\eta=1,4/3,2$), if it is due to differential rotation between the GW-emitting quadrupole and the rigid crust~\cite{SieniawskaJones2022}. However the scenario $\vert f(t)- \eta f_{\star}(t) \vert \sim f_{\star}(t)$ is also conceivable, for example if $r$-mode emission occurs far from $4f_{\star}/3$ due to complicated microphysics \cite{CarideInta2019}. The HMM runs fast, so we elect to play safe and cover a wider range of eventualities in this paper by searching $60\,{\rm Hz} \leq f \leq 500\,{\rm Hz}$.

The search assumes $\dot{f}=0$ throughout, e.g.when evaluating the $\mathcal{J}$-statistic. Table~A1 in Ref.~\cite{PereraDeCesar2019} quotes the measured value, $\dot{f}_{\star}=−1.728350\times10^{-15}\pm8\times10^{-21}$ Hz. It causes a frequency drift $\dot{f}_{\star}T_{\rm obs}$ during the observation which is $10$ times smaller than one frequency bin (see Section~\ref{subsecIV:T_drift_and_fbin}). However this target is known to posses a considerable transverse velocity $\vert \mathbf{v}_{\perp} \vert = 104.71\pm0.95~{\rm km\,s}^{-1}$ \cite{DellerVerbiest2008}. It is natural to ask whether corrections to $\dot{f}_{\star}$, i.e. the Shklovskii effect \cite{Shklovskii1970}, need to be included in the analysis. We explore the effect of  proper motion in the $\mathcal{F}$- and hence ${\cal J}$-statistics in Appendix \ref{AppendixA}, while Appendix~\ref{AppendixB} explains how the Shklovskii effect is naturally included in the $\mathcal{F}$- and hence ${\cal J}$-statistics in their existing implementations in the LIGO Scientific Collaboration Algorithm Library (LALSuite) \cite{lalsuite}. The results in both Appendix \ref{AppendixA} and Appendix \ref{AppendixB} confirm that proper motion  corrections are negligible for our search. As such, we approximate $\alpha$ and $\delta$ as constant throughout the search and persist with $\dot{f}=0$.

\begin{table*}
\caption{\label{tab:Target_Params} Electromagnetically measured source parameters, expressed as the central value and the 1-$\sigma$ uncertainty in parenthesis. All electromagnetically constrained parameter ranges are taken from Ref.~\cite{PereraDeCesar2019}. The time of ascension $T_{\rm asc}$ stands for the  value in \cite{PereraDeCesar2019} propagated to the start of O3, as described in Section \ref{subsecIII:target_params}.}
\begin{ruledtabular}
\begin{tabular}{lccccc}
 Parameter & Symbol & Parameter value & Units \\ \hline
 Right ascension& $\alpha$ & $04\,\text{h}\,37\,\text{m}\,15.9125330(5)\,\text{s}$ & J2000 \\
 Declination& $\delta$ & $-47^{\circ}15'09''.208600(5)$  & J2000\\
 Orbital inclination angle & $\iota$ & $137.51(2)$  & Degrees \\
 Projected semi-major axis & $a_{0}$ & $3.36672001(5)$ & $\text{lt-s}$ \\
 Orbital period& $P$ & $496\,026.357(26)$ & $\text{s}$ \\
 GPS time of ascension & $T_{\rm asc}$ & $1\,265\,652\,972(55)$ & $\text{s}$  \\
 Spin-frequency& $f_{\star}$ & $173.6879457375201(9)$ & $\text{Hz}$  \\
\end{tabular}
\end{ruledtabular}
\end{table*}

\section{Search Implementation}
\label{sectionIV:Search_implementation}

This section presents the practical details of the search. Section~\ref{subsecIV:T_drift_and_fbin} justifies our selection of $T_{\rm drift}$, and its implications for the search.  Section~\ref{subsecIV:Number_of_orb_templates} defines the number and spacing of the orbital parameter templates. Section~\ref{subsecIV:Detection_threshold_and_search_workflow} briefly explains the procedure to set a detection threshold in terms of a false alarm probability. Section~\ref{subsecIV:O3_data} describes the data used for the search.

\subsection{$T_{\rm drift}$ and frequency binning}
\label{subsecIV:T_drift_and_fbin}

The coherence time, $T_{\rm drift}$, is a crucial component of the search as it implicitly defines the proposed signal model as described in Section~\ref{sectionII:HMM_Algorithm}. Shorter coherence times, holding $T_{\rm obs}$ constant, create more segments, so that the signal is free to wander more in the same observation time. The reverse is true for longer coherence times. The frequency resolution, given by the size of the frequency bins $\Delta f$, is set by the coherence time through $\Delta f = 1/(2 T_{\rm drift})$. Using $\Delta f$ and $T_{\rm drift}$, the maximum absolute frequency derivative, $\vert \dot{f} \vert_{\rm max}$, which keeps the signal within a frequency bin during a block of duration $T_{\rm drift}$, is calculated as $\vert \dot{f} \vert_{\rm max} = \Delta f / T_{\rm drift}$. 

In the absence of a GW detection from an MSP to date, there is no way to predict $T_{\rm drift}$ for the GW-emitting quadrupole. Instead we make a plausible estimate by analogy with the rotational irregularities observed in $f_{\star}(t)$ through radio timing experiments, such as timing noise \cite{CordesHelfand1980,CordesGreenstein1981} and glitches \cite{AlparCheng1988}. The typical time-scale associated with deviations from the long-term secular evolution of a rotating NS is typically $\sim10$-$20$ days, estimated by auto-correlating pulse times of arrival \cite{PriceLink2012}. This time-scale is consistent with that observed in post-glitch recoveries \cite{vanEysdenMelatos2008, LyneShemar2000, McCullochHamilton1990, WongBacker2001}. In this paper we bracket the above estimates by conducting two analysis with $T_{\rm drift}^{({\rm i})}=10\,{\rm days}$ and $T_{\rm drift}^{({\rm ii})}=30\,{\rm days}$.

The \Ti-analysis matches the value used for the coherence time in previous HMM searches, allowing for direct comparisons with published results \cite{TheLIGOScientificCollaborationtheVirgoCollaboration2022,TheLIGOScientificCollaborationtheVirgoCollaboration2021,AbbottAbbott2019a,AbbottAbbott2017,MiddletonClearwater2020}. The frequency resolution is $\Delta f^{({\rm i})}=5.787037\times10^{-7}$Hz. To assist with data handling, the search is partitioned into sub-bands. Each sub-band contains $N^{({\rm i})}_{f}=2^{20}$ frequency bins spanning a band of width $\Delta f_{\rm sub}^{({\rm i})}=N^{({\rm i})}_{f}\Delta f^{({\rm i})}=0.6068148$ Hz. We choose a power of two for the number of frequency bins as this accelerates the Fourier transform in the $\mathcal{J}$-statistic calculation \cite{SuvorovaClearwater2017,AbbottAbbott2019a}. The number of total sub-bands to consider in this analysis is $N^{({\rm i})}_{\rm sub}={\rm ceil}\left[(500-60)\,{\rm Hz}/\Delta f^{({\rm i})}_{\rm sub} \right]=725$. For O3 the number of contiguous segments, using \Ti, is $N_{T}^{({\rm i})}=36$. The maximum absolute frequency derivative, for this analysis, is $\vert \dot{f}^{({\rm i})} \vert_{\rm max} \approx 6.7\times10^{-13}\,{\rm Hz\,s}^{-1}$, which is greater than the measured long-term secular value for $\dot{f}_{\star}$ in Table A1 of Ref.~\cite{PereraDeCesar2019}. 

The \Tii-analysis uses $\Delta f^{({\rm ii})}=1.929012\times10^{-7}$Hz, $N_{f}^{({\rm ii})}=2^{21}$, $\Delta f^{({\rm ii})}_{\rm sub}=0.4045431$, and $N_{\rm sub}^{({\rm ii})}=1088$ sub-bands. The number of contiguous segments, using \Tii, is $N_{T}^{({\rm ii})}=12$. The maximum absolute frequency derivative consistent with \Tii, $\vert \dot{f}^{({\rm i})} \vert_{\rm max} \approx 7.5\times10^{-14}\,{\rm Hz\,s}^{-1}$, is one order of magnitude above the electromagnetically measured $\dot{f}_{\star}$ \cite{PereraDeCesar2019}. 

For convenience, the main parameters discussed in this section are summarized in Table~\ref{tab:diff_params_Tdrifts}.

\begin{table}
\caption{Search ranges and minimum and maximum number of grid points. All other parameters, including $a_{0}$ and $P$, take the central values in Table~\ref{tab:Target_Params} in every template. The template ranges are the same for \Ti and \Tii.}
\label{tab:diff_params_Tdrifts} 
\begin{ruledtabular}
\begin{tabular}{lcc}
Parameter  &  Minimum & Maximum   \\ \hline
$f$ / Hz &  $60$ & $500$  \\
$T_{\rm asc}$ / s &  $1\,265\,652\,809$ & $1\,265\,653\,136$  \\ 
Grid points & $6$ & $49$ \\
\end{tabular}
\end{ruledtabular}
\end{table}

\subsection{Number and placing of orbital templates}
\label{subsecIV:Number_of_orb_templates}

We cover the orbital parameter space through a rectangular grid spanning $(a_{0} \pm 3\sigma_{a_{0}}, T_{\rm asc}\pm 3\sigma_{\rm T_{\rm asc}}, P\pm3\sigma_{P})$. The number of grid points depends on a user-selected maximum mismatch $\mu_{\rm max}$ \cite{LeaciPrix2015}, which is the fractional loss in signal-to-noise ratio between the search using the true parameters and the search using the nearest grid point. We use $\mu_{\rm max}=0.1$ in this paper, following many previous analyses \cite{TheLIGOScientificCollaborationtheVirgoCollaboration2022,TheLIGOScientificCollaborationtheVirgoCollaboration2021,AbbottAbbott2019a,AbbottAbbott2017,MiddletonClearwater2020}. The number of grid points along each parameter axis is calculated using Eq.~(71) of Ref.~\cite{LeaciPrix2015}, viz.

\begin{equation}
    N_{a_{0}} = \ceil[\Bigg]{ 3 \sqrt{2} \mu_{\rm max}^{-1/2} f \sigma_{a_{0}} },
    \label{eq:Numbera0}
\end{equation}

\begin{equation}
    N_{T_{\rm asc}} = \ceil[\Bigg]{ 6 \pi^{2} \sqrt{2} \mu_{\rm max}^{-1/2} f a_{0} P^{-1} \sigma_{T_{\rm asc}}},
    \label{eq:NumberTasc}
\end{equation}

\begin{equation}
    N_{P} = \ceil[\Bigg]{\pi^{2}\sqrt{6}\mu_{\rm max}^{-1/2}f a_{0} T_{\text{drift}} N_{T}P^{-2}\sigma_{P}}.
    \label{eq:NumberP}
\end{equation}

The uncertainties in Table~\ref{tab:Target_Params}, together with Eq.~(\ref{eq:Numbera0}) and Eq.~(\ref{eq:NumberP}), imply $N_{a_{0}}=N_{P}=1$ in every sub-band. As Eq.~(\ref{eq:NumberTasc}) depends on $f$, and one has $3\sigma_{T_{\rm asc}}=165$ s (see Table II), $N_{T_{\rm asc}}$ varies for each sub-band within the range $6 \leq N_{T_{\rm asc}} \leq 49$. In total the search covers $\sim10^{4}$ templates, in stark contrast with other searches, such as O3 Scorpius X-1, for which no radio ephemeris exists, and $>10^{9}$ templates are needed \cite{TheLIGOScientificCollaborationtheVirgoCollaboration2022}.

\subsection{Detection threshold}
\label{subsecIV:Detection_threshold_and_search_workflow}

The output of the HMM pipeline, for a given sub-band, is the optimal path and its associated log likelihood, $\rm max(\mathcal{L})$, after scanning over all possible $(a_{0},T_{\rm asc},P)$ triads. The reader is directed to Figure 2 of Ref.~\cite{TheLIGOScientificCollaborationtheVirgoCollaboration2022} for a thorough explanation of the HMM workflow.  

A sub-band is flagged as containing a candidate, when $\rm max(\mathcal{L})$ exceeds a threshold, $\mathcal{L}_{\rm th}$, consistent with a user-selected false alarm probability. We estimate the distribution of $\rm max(\mathcal{L})$ in pure noise, by creating synthetic Gaussian data using the \texttt{lalapp\_Makefakedata\_v5} program in LALSuite, the parameter values found in Table~\ref{tab:Target_Params} and $T_{\rm obs}$. We note this step is done for both \Ti and \Tii. 

The false alarm probability, $\alpha_{N_{\rm tot}}$, of a sub-band with $N_{\rm tot}$ log likelihoods, is a function of the false alarm probability in a single terminating frequency bin per orbital template, $\alpha$, viz.

\begin{equation}
    \alpha_{N_{\rm tot}}=1-(1-\alpha)^{N_{\rm tot}}.
\end{equation}

\noindent We choose to keep the number of candidates manageable for follow-up studies, and to facilitate comparison with many previous HMM searches, by setting $\alpha_{N_{tot}} = 0.01$ \cite{TheLIGOScientificCollaborationtheVirgoCollaboration2022,AbbottAbbott2019a,AbbottAbbott2017}. For a detailed explanation of the threshold procedure, involving simulations with Gaussian data, see Appendix A in Ref.~\cite{TheLIGOScientificCollaborationtheVirgoCollaboration2021}.

\subsection{O3 data}
\label{subsecIV:O3_data}

The search uses all the O3 dataset, collected by the LIGO Hanford and Livingston detectors, dated from April 1, 2019 15:00 UTC to March 27, 2020, 17:00 UTC. We do not include data from the Virgo detectors, given their lower sensitivity compared to the two LIGO observatories in the $60\,{\rm Hz}-500$ Hz band \cite{DavisAreeda2021}. The O3 dataset is divided in two, namely O3a, which spans 1 April 2019 to 1 October 2019, and O3b, which spans 1 November 2019 to 26 March 2020. O3a is followed by a month-long commissioning break. 

Short Fourier transforms (SFTs) are created from the ``CO1 calibrated self-gated 60 Hz subtracted" dataset, which uses the procedure in Ref.~\cite{CalibrationO3} to remove any loud glitches and subtracts the 60 Hz noise line via the procedure described in Ref.~\cite{VajenteHuang2020}. The number of segments affected by the month-long commissioning break depends on $T_{\rm drift}$. Any segment with no SFTs is replaced by a ``fake" segment with uniform log likelihood across all frequency bins, to accommodate spin wandering during the commissioning break. For \Ti, two segments are replaced. For \Tii only one segment is replaced. 

\subsection{Vetoes}
\label{subsecIV:Vetoes}
Every candidate is subjected to a hierarchy of vetoes to differentiate between non-Gaussian artifacts and  astrophysical signals. In this paper, we follow previous works \cite{AbbottAbbott2017,AbbottAbbott2019a,MiddletonClearwater2020,TheLIGOScientificCollaborationtheVirgoCollaboration2022,TheLIGOScientificCollaborationtheVirgoCollaboration2021}, and use the known lines veto described in Section \ref{subsecIV:Vetoes_KL}, and the single interferometer (IFO) veto detailed in Section \ref{subsecIV:Vetoes_SIFO}.

\subsubsection{Known lines}
\label{subsecIV:Vetoes_KL}

Narrowband noise artifacts in the IFO can mimic an astrophysical signal, artificially increasing the $\mathcal{F}$-statistic in certain sub-bands. These noise features can be created by a multitude of causes, including the IFO suspension system or the electricity grid \cite{CovasEffler2018}. We veto any candidate whose optimal frequency path $f(t)$ satisfies 

\begin{equation}
    \vert f(t) - f_{\rm line} \vert < 2 \pi a_{0} f_{\rm line}/P , 
    \label{Eq:subsecIV_vetoes_KL}
\end{equation}

\noindent for any epoch $t$ of the search. In Eq.~(\ref{Eq:subsecIV_vetoes_KL}) $f_{\rm line}$ is the frequency of the closest noise line found in the vetted list in Ref.~\cite{O3lines}.

\subsubsection{Single IFO}
\label{subsecIV:Vetoes_SIFO}

Astrophysical signals are likely to be weak enough to need both IFOs for detection or strong enough to appear simultaneously in both detectors, given the detectors comparable sensitivity throughout the frequency band. In contrast, the opposite is true for non-astrophysical signals caused by instrumental artifacts in an IFO. 

Let \maxL$_{a}$ and \maxL$_{b}$, with \maxL$_{a} >~$\maxL$_{b}$, denote the max log likelihoods recovered for IFO $a$ and IFO $b$, separately and respectively. We use \maxL$_{\cup}$ to denote the original candidate's max log likelihood, with associated optimal path $f_{\cup}(t)$. The four possible outcomes are:

\begin{enumerate}
    \item If one finds \maxL$_{a} >\;$\maxL$_{\cup}$ and \maxL$_{b} < \mathcal{L}_{\rm th}$, and the optimal path for \maxL$_{a}$ viz. $f_{a}(t)$, satisfies
    \begin{equation}
        \vert f_{\cup}(t) - f_{a}(t) \vert < 2 \pi a_{0} f_{\cup}/P ,
        \label{Eq:subsecIV_vetoes_SIFO}
    \end{equation}
    \noindent for any search epoch, then the candidate's behaviour is consistent with an instrumental artifact in IFO $a$. This candidate is vetoed.
    \item If one finds \maxL$_{a} >\;$\maxL$_{\cup}$ and \maxL$_{b} < \mathcal{L}_{\rm th}$, but Eq.~(\ref{Eq:subsecIV_vetoes_SIFO}) is not satisfied, then the candidate could be a faint astrophysical signal. The candidate is saved for post-processing.
    \item If one finds \maxL$_{a} > \mathcal{L}_{\rm th}$ and \maxL$_{b} > \mathcal{L}_{\rm th}$, then the candidate could be a strong astrophysical signal, or a common noise artifact in both detectors. The candidate is saved for post-processing.
    \item If one finds \maxL$_{a} < \mathcal{L}_{\cup}$ and \maxL$_{b} < \mathcal{L}_{\rm th}$, then the candidate could be a weak astrophysical signal. The candidate is saved for post-processing.
\end{enumerate}

\section{Analysis}
\label{SectionV:Analysis_Results}

The outputs of the search and veto procedures are presented for \Ti~and \Tii~in Sections \ref{subsecV:Ti} and \ref{subsecV:Tii} respectively.

\begin{figure*}[ht]
    \centering
    \includegraphics[scale=0.45]{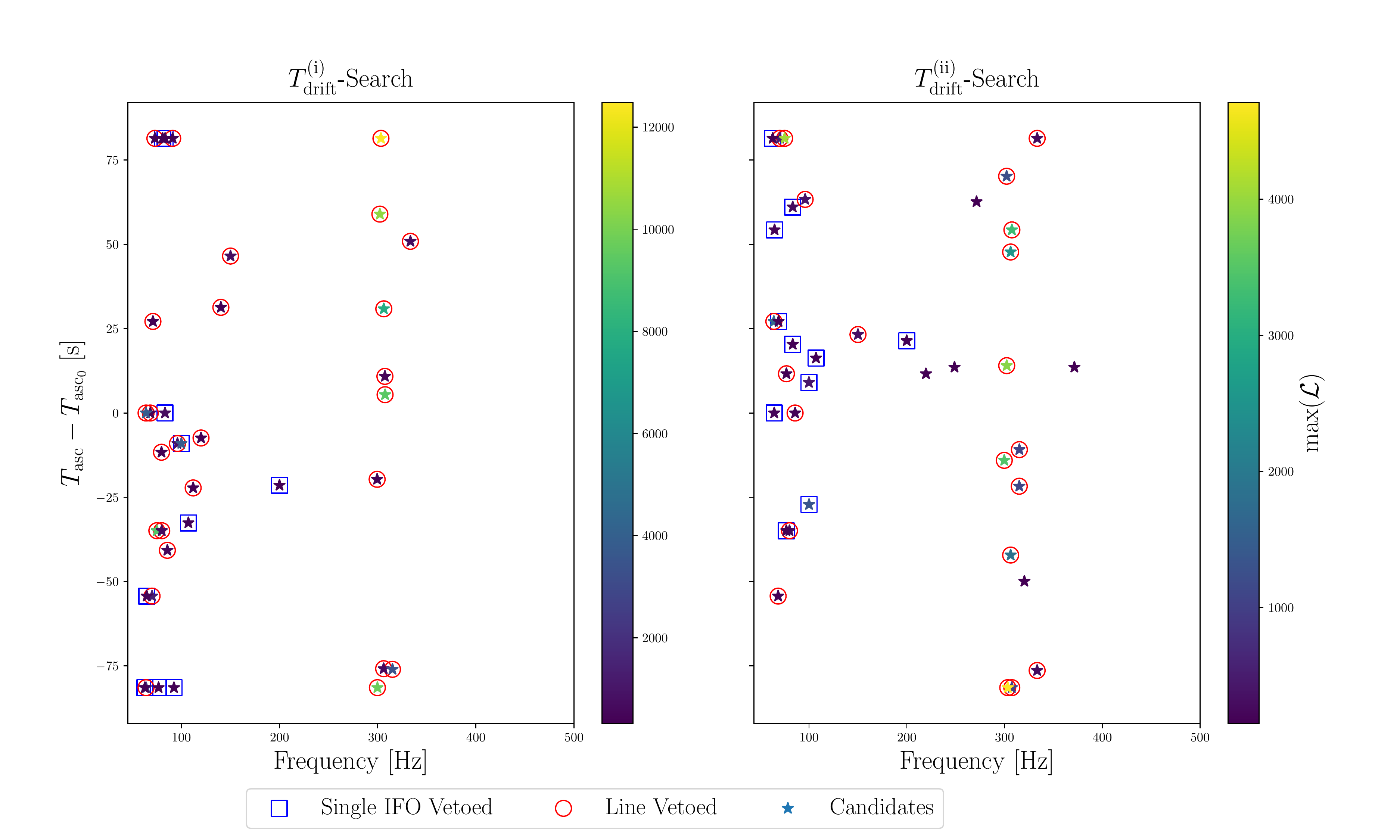}
    \caption{Candidates (denoted by stars) plotted as a function of their frequency bin (horizontal axis) and the offset from the central time of ascension $T_{\rm asc}-T_{\rm asc_{0}}$ (vertical axis), for the \Ti~(left panel) and the \Tii~(right panel) analyses. The color scale represents ${\rm max}(\mathcal{L})$. Stars decorated with purple squares or red circles are eliminated by the single IFO or known lines veto, respectively. Undecorated stars survive both vetoes and are followed up further.}   
    \label{fig:TascSpaceGraph}
\end{figure*}

\subsection{\Ti~candidates}
\label{subsecV:Ti}

We obtain $36$ candidates with \maxL~$>\mathcal{L}_{\rm th}$. After applying the hierarchy of vetoes none of the candidates survives. The known lines veto eliminates $26$ candidates, while the single IFO veto eliminates the remaining $10$. Table~\ref{tab:Candidates_Ti} summarizes the outcome of the vetoes in the 36 sub-bands containing a candidate.

The results of this analysis are plotted in the left panel of Figure~\ref{fig:TascSpaceGraph}. The vertical axis shows the offset of the candidate's $T_{\rm asc}$ value from the central value found in Table~\ref{tab:Target_Params} (denoted as $T_{\rm asc,0}$ in Figure~\ref{fig:TascSpaceGraph}). The horizontal axis corresponds to the terminating frequency bin of the optimal path, $q^{*}(t_{N_{T}})$, of the candidate.

\begin{table*}[tbp]
\caption{\label{tab:Candidates_Ti} Veto outcomes for the \Ti~analysis. The first column is the starting frequency of the sub-band containing the candidate. The second column is the \maxL~of the candidate. The two last column present the results of the vetoes presented in Section~\ref{subsecIV:Vetoes}. \checkmark~indicates the candidate survives the veto, while \textbf{X} denotes the opposite.}
\begin{ruledtabular}
\begin{tabular}{cccc}%c
Sub-band $(\text{Hz})$ & $\rm max(\mathcal{L})/10^{3}$ & Known lines veto & Single IFO veto \\ \hline 
62.49 & 0.39 & \checkmark & \textbf{X} \\
63.09 & 3.79 & \textbf{X} & \textbf{-} \\
63.70 & 2.38 & \textbf{X} & \textbf{-} \\
64.31 & 0.47 & \checkmark & \textbf{X} \\
67.95 & 0.44 & \textbf{X} & \textbf{-} \\
69.77 & 1.39 & \textbf{X} & \textbf{-} \\
70.38 & 0.32 & \textbf{X} & \textbf{-} \\
72.20 & 0.37 & \textbf{X} & \textbf{-} \\
74.62 & 9.22 & \textbf{X} & \textbf{-} \\
76.44 & 0.36 & \checkmark & \textbf{X} \\
78.87 & 0.35 & \textbf{X} & \textbf{-} \\
79.48 & 0.69 & \textbf{X} & \textbf{-} \\
80.08 & 0.32 & \checkmark & \textbf{X} \\
82.51 & 0.79 & \checkmark & \textbf{X} \\
83.12 & 0.41 & \checkmark & \textbf{X} \\
84.94 & 0.53 & \textbf{X} & \textbf{-} \\
90.40 & 0.32 & \textbf{X} & \textbf{-} \\
91.61 & 0.32 & \checkmark & \textbf{X} \\
95.25 & 0.90 & \textbf{X} & \textbf{-} \\
99.50 & 3.76 & \checkmark & \textbf{X} \\
106.78 & 0.39 & \checkmark & \textbf{X} \\
111.64 & 0.35 & \textbf{X} & \textbf{-} \\
119.53 & 0.32 & \textbf{X} & \textbf{-} \\
139.55 & 0.38 & \textbf{X} & \textbf{-} \\
149.26 & 0.76 & \textbf{X} & \textbf{-} \\
199.63 & 0.40 & \checkmark & \textbf{X} \\
298.54 & 0.40 & \textbf{X} & \textbf{-} \\
299.14 & 9.75 & \textbf{X} & \textbf{-} \\
301.57 & 10.48 & \textbf{X} & \textbf{-} \\
302.79 & 12.49 & \textbf{X} & \textbf{-} \\
305.21 & 0.62 & \textbf{X} & \textbf{-} \\
305.82 & 7.79 & \textbf{X} & \textbf{-} \\
306.43 & 0.55 & \textbf{X} & \textbf{-} \\
307.03 & 9.35 & \textbf{X} & \textbf{-} \\
314.31 & 3.49 & \textbf{X} & \textbf{-} \\
332.52 & 0.66 & \textbf{X} & \textbf{-} \\ \hline
Total: 36 &  &	 &	  \\
\end{tabular}
\end{ruledtabular}
\end{table*}

\subsection{\Tii~candidates}
\label{subsecV:Tii}

The \Tii~analysis produces 37 candidates above the threshold. Out of the $37$ candidates, the known lines veto eliminates $22$ and the single IFO eliminates $10$. All of the surviving candidates belong to outcome $4$ in Section~\ref{subsecIV:Vetoes_SIFO}. Table~\ref{tab:Candidates_Tii} presents the results of the veto procedure for the candidate's sub-band. The results of the analysis are plotted in the right side panel of Figure~\ref{fig:TascSpaceGraph}.  Five candidates survive the hierarchy of vetoes defined in Section~\ref{subsecIV:Vetoes}. Information about the surviving candidates is summarized in Table~\ref{tab:params_Survivors} in Appendix~\ref{AppendixC}. We note that the number of survivors is consistent with the expected number of false alarms, viz. $0.01\,N^{({\rm ii})}_{\rm sub}\approx11$. 

Among the five survivors, two occur at the same value of $T_{\rm asc}$, viz. $T_{\rm asc}= 1265652985.7228$ s. Perhaps coincidentally, the terminating frequencies of their optimal HMM paths are in the ratio $1.493 \approx 3/2$. One may seek conceivably to interpret these survivors as the first and second harmonics of a fundamental mode at $\approx 124.3 \, {\rm Hz}$, but no candidate with ${\cal L} > {\cal L}_{\rm th}$ or known noise line \cite{O3lines} occurs near the latter frequency. More broadly, the frequency ratios of the five survivors are not simple fractions, as might occur with a harmonic series of instrumental or astrophysical origin.

In a similar vein, one may ask whether the frequency differences between the five survivors are simple rational multiples of an astrophysical frequency (such as $f_{\star}$) or an instrumental frequency (such as a noise line). Interestingly, the first, third, fourth, and fifth survivors in Table~\ref{tab:Candidates_Tii} are separated by integer multiples of $\approx 50\,{\rm Hz}$, and instrumental noise lines exist at $f_{\rm line,1} = 99.88 \, {\rm Hz}$ and $f_{\rm line,2}= 150.05 \, {\rm Hz}$ which feature in the vetoes in Section~\ref{subsecIV:Vetoes}. Arguably, therefore, at most one (if any, and not necessarily the first) of these survivors could be astrophysical. However, it is important to note that the frequency differences between the first, third, fourth, and fifth survivors differ from simple rational multiples of $f_{\rm line,1}$ and $f_{\rm line,2}$ by up to $2\Delta f_{\rm sub}^{\rm (ii)} = 0.80908642 \, {\rm Hz}$, so the pattern may be pure coincidence. The frequency differences involving the second survivor at $248.80 \, {\rm Hz}$ are not simple rational multiples of any obvious frequency, and the second survivor shares the same $T_{\rm asc}$ as the fifth survivor at $371.42\,{\rm Hz}$. Hence one or both of the survivors at $248.80 \, {\rm Hz}$ and $371.42\, {\rm Hz}$ are consistent with astrophysical or instrumental origins; all combinations remain viable without further information. The second and fifth survivors differ by $7f_\ast/10$ to within $\approx 1\%$.

We follow up the survivors by plotting how the detection statistic varies with the sky position of the search. We  calculate $\mathcal{L}$ for different $\alpha$ and $\delta$ for a uniform $(\alpha,\delta)$-grid covering $0.23\,{\rm arcmin}^{2}$ around the source position. The aim of this test is to explore if the effective ``point spread function" (by analogy with electromagnetic telescopes) generated by the candidate, as described through $\mathcal{L}(\alpha,\delta)$, follows the expected behaviour of an astrophysical signal~\cite{TheLIGOScientificCollaborationtheVirgoCollaboration2021,JonesSun2022}. 

\begin{figure}[ht]
    \includegraphics[scale=0.5, center]{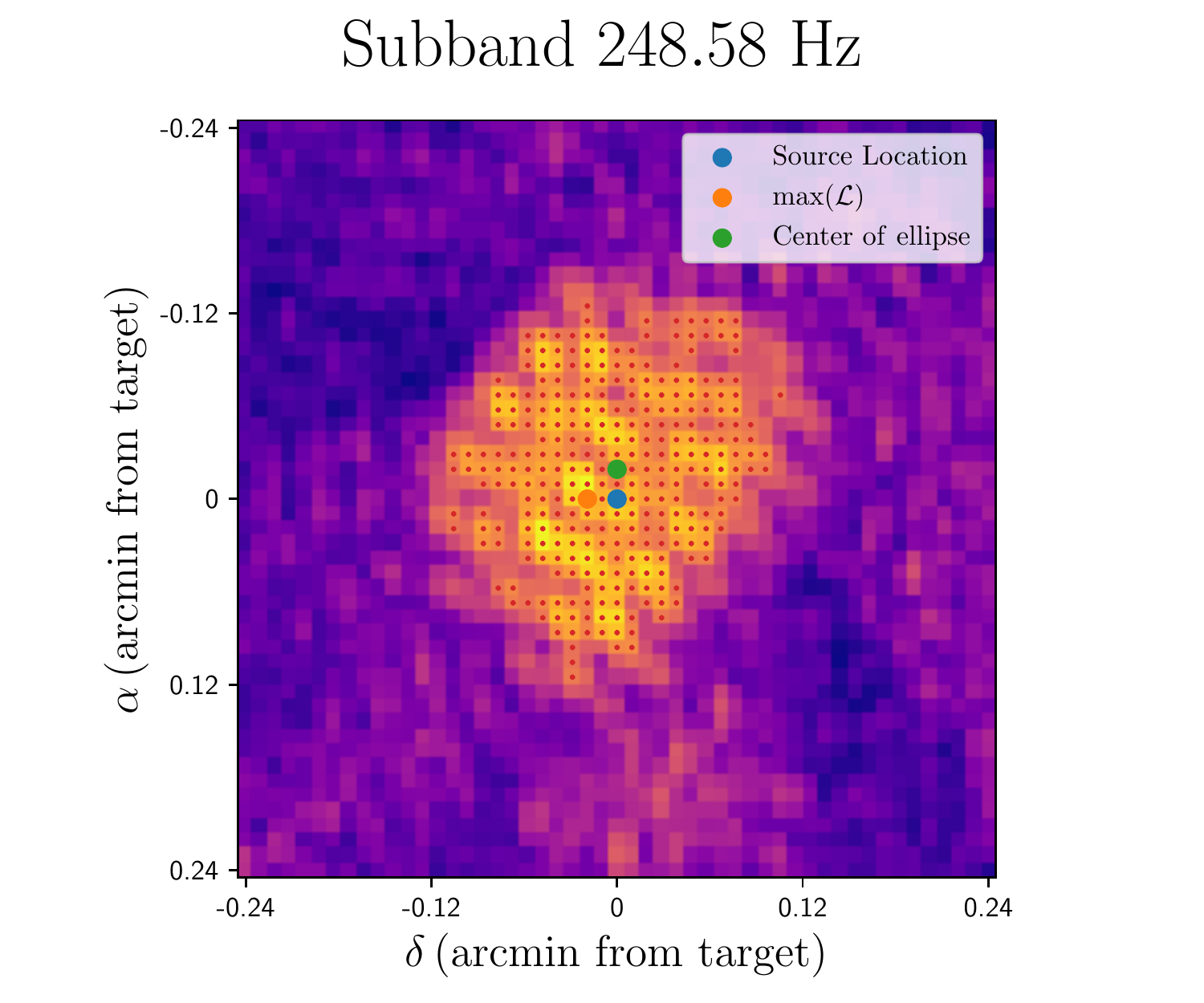}
    \caption{Sky-map for the $248.58$ Hz sub-band candidate. The source location (blue dot), maximum log likelihood location~[\maxL~; orange dot],  and center of the ellipse (green dot) are marked. The center of the ellipse is calculated by averaging over all pixels that satisfy $\mathcal{L} > \mathcal{L}_{\rm th}$ (marked by red dots). In general the three dots should coincide for an astrophysical signal; here they are separated by $ 0.006\,{\rm arcsec}$. }   
    \label{fig:Example_SkyMap_Focii}
\end{figure}

The results of the sky position study take the form of sky maps such as the one displayed in Figure~\ref{fig:Example_SkyMap_Focii} for the survivor in the sub-band at $248.58$ Hz. The sky maps for the other four survivors are presented in Appendix~\ref{AppendixC}. In general an astrophysical candidate's $\mathcal{L}$ should diminish when pointing away from the source, forming elliptical contours of constant $\mathcal{L}(\alpha,\delta)$. In Figure~\ref{fig:Example_SkyMap_Focii}, the source location, maximum log likelihood across the map [\maxL] and the center of the ellipse are marked by a blue, orange, and green dot, respectively. The center of the ellipse is calculated by averaging the position of pixels satisfying $\mathcal{L} > \mathcal{L}_{\rm th}$ (marked by red dots in Figure~\ref{fig:Example_SkyMap_Focii}).  For a strong enough astrophysical signal all three locations should coincide. None of the surviving candidates display such coincidence either in Figure~\ref{fig:Example_SkyMap_Focii} or in Appendix~\ref{AppendixC}. The closest three-way coincidence is within $0.006$ arcsec for the survivor at $248.58$ Hz. All surviving candidates are weak, i.e. they are barely above $\mathcal{L}_{\rm th}$ (see Table~\ref{tab:params_Survivors}). This translates into small ellipses with area $\sim 207~{\rm arcsec}^{2}$ around each candidate, in contrast to the $\sim 4~{\rm arcmin}^{2}$ ellipse in Figure 6 in Ref.~\cite{TheLIGOScientificCollaborationtheVirgoCollaboration2021}.

\begin{table*}[tbp]
\caption{\label{tab:Candidates_Tii} Veto outcomes for the \Tii~analysis, arranged as in Table~\ref{tab:Candidates_Ti}. In the last column H and L are the Hanford-only and Livingston-only \maxL~achieved by the candidate.}
\begin{ruledtabular}
\begin{tabular}{cccc}%c
Sub-band $(\text{Hz})$ & $\rm max(\mathcal{L})/10^{3}$ & Known lines veto & Single IFO veto \\ \hline 
62.49 & 0.15 & \checkmark & \textbf{X} \\
63.70 & 1.64 & \textbf{X} & \textbf{-} \\
64.10 & 0.20 & \checkmark & \textbf{X} \\
64.51 & 0.19 & \checkmark & \textbf{X} \\
68.15 & 0.19 & \textbf{X} & \textbf{-} \\
68.55 & 0.17 & \checkmark & \textbf{X} \\
69.77 & 0.52 & \textbf{X} & \textbf{-} \\
74.62 & 4.08 & \textbf{X} & \textbf{-} \\
76.24 & 0.15 & \checkmark & \textbf{X} \\
76.65 & 0.15 & \textbf{X} & \textbf{-} \\
79.48 & 0.27 & \textbf{X} & \textbf{-} \\
82.71 & 0.27 & \checkmark & \textbf{X} \\
83.12 & 0.18 & \checkmark & \textbf{X} \\
85.14 & 0.19 & \textbf{X} & \textbf{-} \\
95.66 & 0.39 & \textbf{X} & \textbf{-} \\
99.30 & 0.42 & \checkmark & \textbf{X} \\
99.70 & 1.36 & \checkmark & \textbf{X} \\
106.58 & 0.15 & \checkmark & \textbf{X} \\
149.46 & 0.34 & \textbf{X} & \textbf{-} \\
199.63 & 0.16 & \checkmark & \textbf{X} \\
219.04 & 0.15 & \checkmark & \checkmark; H: 120.81, L: 124.15 \\
248.58 & 0.15 & \checkmark & \checkmark; H: 124.06, L: 127.83 \\
270.83 & 0.15 & \checkmark & \checkmark; H: 120.07, L: 123.50 \\
299.14 & 3.45 & \textbf{X} & \textbf{-} \\
301.57 & 1.25 & \textbf{X} & \textbf{-} \\
301.98 & 3.91 & \textbf{X} & \textbf{-} \\
302.79 & 4.71 & \textbf{X} & \textbf{-} \\
305.62 & 2.52 & \textbf{X} & \textbf{-} \\
306.02 & 1.86 & \textbf{X} & \textbf{-} \\
306.83 & 1.02 & \textbf{X} & \textbf{-} \\
307.24 & 3.28 & \textbf{X} & \textbf{-} \\
314.52 & 1.13 & \textbf{X} & \textbf{-} \\
314.92 & 1.06 & \textbf{X} & \textbf{-} \\
319.78 & 0.15 & \checkmark & \checkmark; H: 123.07, L: 118.09 \\
332.72 & 0.28 & \textbf{X} & \textbf{-} \\
333.13 & 0.26 & \textbf{X} & \textbf{-} \\
371.15 & 0.15 & \checkmark & \checkmark; H: 123.21, L: 124.17 \\ \hline
Total: 37 &  &	 &	  \\
\end{tabular}
\end{ruledtabular}
\end{table*}

\section{Conclusions}
\label{SectionVI:Conclusions}

In this paper we search for continuous GWs from \target~in LIGO O3 data, using a HMM combined with the maximum-likelihood $\mathcal{J}$-statistic and a template grid for the time of ascension $T_{\rm asc}$. As it is eminently possible to have $f(t)\neq f_{\star}(t)$, we search the band $60\,{\rm Hz}\leq f \leq 500\,{\rm Hz}$ to play safe and cover a range of possibilities within the most sensitive LIGO band, generalizing previous narrowband searches at simple rational multiples of $f_{\star}$ \cite{AbbottAbbott2019b,AbbottAbbott2020,TheLIGOScientificCollaborationtheVirgoCollaboration2021a}. As the underlying wandering time-scale of the GW-emitting quadruopole is unknown \cite{CordesHelfand1980,CordesGreenstein1981,AlparCheng1988,PriceLink2012,vanEysdenMelatos2008, LyneShemar2000, McCullochHamilton1990, WongBacker2001}, we conduct two distinct analysis with $T_{\rm drift}^{({\rm i})}=10$ days and $T_{\rm drift}^{({\rm ii})}=30$ days. Each analysis uses a detection threshold, $\mathcal{L}_{\rm th}$, with a false alarm probability of $\alpha_{N_{\rm tot}}=0.01$ per sub-band, set via Monte-Carlo simulations. The number of sub-bands is $725$ and $1088$ for the \Ti~and \Tii~analyses, respectively.

Any sub-band with an optimal path satisfying \maxL$>\mathcal{L}_{\rm th}$ is tested by a hierarchy of vetoes. No candidate survives the veto procedure for the \Ti~analysis, while five candidates survive the \Tii~analysis, two of which share the same $T_{\rm asc}$ and are in the frequency ratio $1.493$. The number of survivors is consistent with the expected number of false alarms. Sky maps of ${\cal L}(\alpha,\delta)$ versus search position do not reveal clear signatures of instrumental artifacts or an astrophysical origin. All five survivors appear in sub-bands not covered by previous analyses \cite{AbbottAbbott2019b,AbbottAbbott2020,TheLIGOScientificCollaborationtheVirgoCollaboration2021a}. We record their frequencies and orbital templates in Table~\ref{tab:params_Survivors}. Future searches, including during the upcoming fourth LIGO-Virgo-KAGRA observing run, will shed more light on the nature of the five survivors.

\section{Acknowledgements}

This work used computational resources of the OzSTAR national facility at Swinburne University of Technology. OzSTAR is funded by Swinburne University of Technology and also the National Collaborative Research Infrastructure Strategy (NCRIS). This material is based upon work supported by NSF's LIGO Laboratory which is a major facility fully funded by the National Science Foundation. This work has been assigned LIGO document number P2200212. 

\appendix

\section{Proper motion in the ${\cal F}$- and ${\cal J}$-statistics}
\label{AppendixA}

The $\mathcal{F}$-statistic (and hence the ${\cal J}$-statistic, which is a linear combination of ${\cal F}$-statistic values) assumes a signal model that represents a quadrupolar GW emitted by a freely precessing axisymmetric rotor \cite{JaranowskiKrolak1998}. The amplitude of the signal depends on the sky position of the source, i.e. $\alpha$ and $\delta$, through the beam-pattern functions $F_{+}(t)$ and $F_{\times}(t)$ (as expressed in Eq.~(10)-(13) in Ref.~\cite{JaranowskiKrolak1998}). The phase of the signal, $\Psi(t)$, depends on the transverse velocity of the source projected onto the plane of the sky, $\mathbf{v}_{\perp}$, as expressed in Eq.~(12) in Ref.~\cite{JaranowskiKrolak1999}. The proper motion of \target~is known to be significant in the context of phase-connected radio pulsar timing experiments \cite{Shklovskii1970,BellBailes1993,BellBailes1995}. It is therefore natural to ask whether it is also significant in the context of HMM GW searches, which also rely on partial phase connection; for example the ${\cal F}$- and ${\cal J}$-statistics assume phase coherence during intervals of length $T_{\rm drift}$, as discussed in Section~\ref{sectionII:HMM_Algorithm}. We explore this issue in this appendix. 

The sky position of the source changes with time to leading order according to 

\begin{align}
    \alpha(t) &= \alpha + \mu_{\alpha} t \label{ApendixA_alphat}, \\
    \delta(t) &= \delta+\mu_{\delta}t \label{ApendixA_deltat},
\end{align}

\noindent where $\alpha$ and $\delta$ are the right ascension and declination at time $t=0$ (taken to be the values in Table~\ref{tab:Target_Params}, dated  $T_{0}$), and $\mu_{\alpha}$ and $\mu_{\delta}$ are the respective components of the source's proper motion. Electromagnetically measured values for $\mu_{\alpha}$ and $\mu_{\delta}$ can be found in Table 1 of Ref.~\cite{DellerVerbiest2008}, namely $\mu_{\alpha}=121.679(52)~{\rm mas\,yr}^{-1}$, and $\mu_{\delta}=-71.820(86)~{\rm mas\,yr}^{-1}$. Setting $t=T_{\rm obs}=360$ days in Eq.~(\ref{ApendixA_alphat}) and Eq.~(\ref{ApendixA_deltat}), we find the maximum angular displacement of the source from its original position is $0.12$ arcsec and $-0.071$ arcsec in $\alpha$ and $\delta$ respectively.
 
In order to double-check that the proper motion has a negligible effect on the ${\cal F}$- and hence ${\cal J}$-statistics, we conduct the following Monte Carlo test. We generate 100 sets of synthetic data, consisting of a \target-like signal plus Gaussian noise. Each simulation takes place in a sub-band of $0.606814~{\rm Hz}$, centered at $f_{\star}$. The injected signal binary parameters, $\{a_{0},T_{\rm asc},P\}_{\rm inj}$, correspond to their central values in Table~\ref{tab:Target_Params}. For each simulation we generate two sets of atoms, the first set follows the proper motion of \target~with $\alpha(t)$ and $\delta(t)$ varying according to (\ref{ApendixA_alphat}) and (\ref{ApendixA_deltat}), while the second set is fixed at the initial sky coordinates, with $\alpha(t)=\alpha$ and $\delta(t)=\delta$. We use the subscript PM to denote the former set of atoms and ZPM for the latter set.  We summarize the result of the simulations in terms of the mean difference between log likelihoods per atom set. We find $\langle \mathcal{L}_{{\rm ZPM}}-\mathcal{L}_{{\rm PM}} \rangle=0.13$ and $\langle \mathcal{L}_{{\rm ZPM}}-\mathcal{L}_{{\rm PM}} \rangle=0.55$ for \Ti~and \Tii~respectively.

Given the small differences between maximum log likelihoods found above, we perform the search with $\alpha(t)=\alpha$ and $\delta(t)=\delta$ fixed in this paper instead of propagating according to (\ref{ApendixA_alphat}) and (\ref{ApendixA_deltat}). However, to be conservative, we reduce the thresholds per sub-band (See Section~\ref{subsecIV:Detection_threshold_and_search_workflow}) by $0.13$ for \Ti, and by $0.55$ for \Tii.    

$\Psi(t)$ is composed of four contributions: (i) the rotation of the star, (ii) Earth's orbital motion around the SSB, (iii) Earth's rotation, and (iv) the proper motion of the star. In this appendix we focus on contribution (iv), denoted by $\Psi_{\rm pm}(t)$, which is given by Eq.~(16) of Ref.~\cite{JaranowskiKrolak1999}, namely

\begin{equation}
    \Psi_{\rm pm}(t) = 2\pi \left[\frac{\mathbf{v}_{\perp}\cdot\mathbf{r}_{\rm ES}(t)}{c D}\right]ft,
    \label{Appendix_A: phase proper motion star}
\end{equation}

\noindent where $\mathbf{r}_{\rm ES}(t)$ is the vector joining the SSB with the center of the Earth. The maximum number of cycles $\Psi_{\rm pm}(t)$ contributes during the full observation of duration $T_{\rm obs}$ is calculated using \target~values for $\vert \mathbf{v}_{\perp} \vert =104.71\pm0.95~{\rm km/s}$ \cite{DellerVerbiest2008} and $D=156.3\pm1.3{\rm~pc}$ \cite{DellerVerbiest2008}, $\vert \mathbf{r}_{\rm ES}(t) \vert=1~{\rm AU}$, ${\rm max}(f)=500~{\rm Hz}$, and $t=T_{\rm obs}$ in Eq.~(\ref{Appendix_A: phase proper motion star}). This yields

\begin{equation}
    \Psi_{\rm pm}^{\rm max}(T_{\rm obs}) = 0.0342 \pi,
    \label{Appendix_A:results_cycles_pm_phase}
\end{equation}

\noindent which is well below the quarter-cycle criterion set in Appendix A of Ref.~\cite{JaranowskiKrolak1999} to ignore a phase contribution. The loss in signal-to-noise ratio is quantified as $1-{\rm FF}$ where ${\rm FF}$ is the fitting factor that measures the offset between a signal and a filter (defined in Eq.~(65) of Ref.~\cite{JaranowskiKrolak1999}). For contribution (iv), we find

\begin{equation}
    1-{\rm FF} \geq 1 - \frac{\sin( 0.0342 \pi)}{ 0.0342 \pi} \approx 0.002,
\end{equation}

\noindent that is, the loss in signal-to-noise ratio does not exceed $0.2\%$. We simplify the search by dropping $\Psi_{\rm pm}(t)$ from $\Psi(t)$.

\section{Shklovskii effect}
\label{AppendixB}

The Shklovksii effect refers to the increase of the observed frequency derivative $\vert \dot{f}_{\star}^{({\rm obs})} \vert$ of a pulsar due to its proper motion \cite{Shklovskii1970}. Pulsars with high transverse velocities appear to accelerate in the reference frame of an Earth-bound observer. For \target, this effect is considerable. Following Ref.~\cite{Shklovskii1970} the kinetic correction due to the Shklovskii effect is given by

\begin{equation}
\dot{f}_{\rm K} = -\frac{\vert {\bf v}_{\perp} \vert^{2}}{cD}f_{\star},
\label{Appendix_B:Shklovskii_effect}
\end{equation}

\noindent where $\dot{f}_{\rm K}$ is the (kinetic) Shklovskii correction to $\dot{f}$. It is well known that the Shklovskii effect is significant in radio pulsar timing experiments targeting \target, because its proper motion is relatively high, and it is located relatively near the Earth $D=156.3\pm1.3{\rm~pc}$ \cite{BellBailes1995,VerbiestBailes2008}. It is therefore natural to ask whether the Shklovskii effect is significant in GW searches too, e.g. when calculating the ${\cal F}$- and ${\cal J}$-statistics. We study this question in this appendix. 

Following Appendix A of Ref.~\cite{JaranowskiKrolak1998}, the phase of the GW signal, in the rest frame of the NS, is given by

\begin{equation}
    \Psi_{\star}(\tau) = \Phi_{0}+2\pi \sum_{k=0}^{s} f_{\star}^{(k)} \frac{\tau^{k+1}}{(k+1)!},
    \label{Eq:GW_phase_NS_RF}
\end{equation}

\noindent where $\tau$ is the proper time in the NS rest frame, and $f_{\star}^{(k)}$ is the $k$th time derivative of $f_{\star}(t)$ evaluated at $\tau=0$. The NS is assumed to move at uniform velocity. Its position vector with respect to the origin of the SSB is given by

\begin{equation}
    \n{r}_{\star}(t)=D\n{n}_{0}+\vert \n{v}_{\star} \vert \n{n}_{v}\left(t+\frac{D}{c}\right),
    \label{Eq:NS_prop_mo_wrt_SSB}
\end{equation}

\noindent with $\n{n}_{o}= {\bf r}_\star(t) / |{\bf r}_\star(t) |$, $\n{v}_{\star}=d {\bf r}_\star(t)/dt$,  and $\n{n}_{v}=\n{v}_{\star}/v_{\star}$. The global time $t$ in Eq.~(\ref{Eq:NS_prop_mo_wrt_SSB}) is the time coordinate in the SSB rest frame.  Eq.~(\ref{Eq:NS_prop_mo_wrt_SSB}) excludes the binary orbit of the NS, because in this appendix we study the effect of proper motion on the ${\cal F}$-statistic for an isolated source. The binary orbit is included later, when constructing the ${\cal J}$-statistic from a linear, Jacobi-Anger combination of ${\cal F}$-statistic values.~\cite{SuvorovaClearwater2017}

The wavefront recorded at the SSB at some time $t$ is originally emitted at  $t'$, with

\begin{equation}
    t=t'+\frac{\vert \n{r}_{\star}(t') \vert}{c}.
    \label{Eq:retarded_times_t_t'}
\end{equation}

\noindent The time $t'$ is related to $\tau$ through special relativistic time dilation, 

\begin{equation}
    \tau = \sqrt{1-\beta_{\star}^{2}}\left( t'+\frac{D}{c}\right),
    \label{Eq:LT_tau_to_t'}
\end{equation}

\noindent with $\beta_{\star} = \vert \n{v}_{\star} \vert/c$. Eq.~(\ref{Eq:LT_tau_to_t'}) assumes $\tau=0$, when the star is located at $\n{r}_{\star}=D\n{n}_{0}$. The GW signal phase at the SSB, $\Psi_{\rm SSB}(t)$, takes the form (see Appendix A of \cite{JaranowskiKrolak1998})

\begin{align}
    \Psi_{\rm SSB}(t) =& \Phi_{0}+2\pi \sum_{k=0}^{s} \frac{f_{\star}^{(k)}}{(k+1)!} (1-\beta^{2}_{\star})^{(k+1)/2}\times \nonumber \\ 
    &\times \left[  t'(t)+\frac{D}{c} \right]^{k+1}, 
    \label{Eq:GW_phase_SSB_RF}
\end{align}

\noindent where $t'(t)$ is the solution of the implicit equation (\ref{Eq:retarded_times_t_t'}) for $t$ fixed:

\begin{align}
    t' = \frac{1}{1-\beta^{2}_{\star}}&\bigg\{t+\frac{D}{c}\beta_{\star} \left[ \beta_{\star}+\n{n}_{0}\cdot\n{n}_{v}\right] \nonumber \\ 
    &-\bigg[\beta^{2}_{\star}t^{2}+2\frac{D}{c}\beta^{2}_{\star}\left( \beta_{\star}+\n{n}_{0}\cdot\n{n}_{v}\right)t \nonumber \\
    &+\frac{D^{2}}{c^{2}}\left(1+\beta_{\star}\n{n}_{0}\cdot\n{n}_{v}\right) \bigg]^{1/2} \bigg\} \label{Eq:time_t'_terms_of_t}.
\end{align}

Upon Taylor expanding $\Psi_{\rm SSB}(t)$ around $t=0$ to quadratic order, using Eqs.~(\ref{Eq:GW_phase_SSB_RF}) and (\ref{Eq:time_t'_terms_of_t}), we find

 \begin{align}
     \frac{\Psi_{\rm SSB}-\Phi_{0}}{2\pi} =& f^{(0)}_{\rm SSB}t \nonumber \\
     &+\bigg\{ f^{(1)}_{\rm SSB} + \frac{[(\n{n}_{0}\cdot\n{n}_{v})^{2}-1]\beta_{\star}^{2}}{(1+\beta_{\star}\n{n}_{0}\cdot\n{n}_{v})^{2}(D/c)}f^{(0)}_{\rm SSB}\bigg\}\frac{t^{2}}{2} \nonumber \\
     &+\mathcal{O}(t^{3}) \label{Eq:phase_SSB_expansion},
 \end{align}

\noindent where 

\begin{equation}
    f^{(k)}_{\rm SSB} = \frac{(1-\beta^{2}_{\star})^{(k+1)/2}}{(1+\beta_{\star}\n{n}_{0}\cdot\n{n}_{v})^{k+1}}f^{(k)}_{\star} \label{Eq:fSSB_terms_fNS}
\end{equation}

\noindent is the Doppler corrected frequency ($k=0$) or its derivatives ($k>0$), as measured in the SSB frame. The reader is directed to Eq.~(119) in Ref.~\cite{JaranowskiKrolak1998} for the complete Taylor expansion of $\Psi_{\rm SSB}(t)$. 

The first term in square brackets in Eq.~(\ref{Eq:phase_SSB_expansion}) equals the frequency derivative observed in the SSB, i.e. $f^{(1)}_{\rm SSB}=\dot{f}_{\star}^{({\rm obs})}$. The second term,  

\begin{equation}
    \Delta f^{(1)}_{\rm SSB} = \frac{[(\n{n}_{0}\cdot\n{n}_{v})^{2}-1]\beta_{\star}^{2}}{(1+\beta_{\star}\n{n}_{0}\cdot\n{n}_{v})^{2}(D/c)}f^{(0)}_{\rm SSB},
    \label{Eq:correction_fdot}
\end{equation}
\noindent is a correction proportional to $f_{\star}$ as observed at the SSB. In \target, one has $\beta_{\star} \approx 3.5\times10^{-4} \ll 1$, which justifies the Taylor expansion

\begin{align}
    \Delta f^{(1)}_{\rm SSB} \approx -\frac{c [1-(\n{n}_{0}\cdot\n{n}_{v})^{2}] f^{(0)}_{\rm SSB}}{D}\beta^{2}_{\star} +\mathcal{O}(\beta^{3}_{\star}).
    \label{Eq:expansion_second_term}
\end{align}

\noindent Writing the transverse velocity as $\n{v}_{\perp}=\n{v}_{\star}-(\n{n}_{0}\cdot\n{v}_{\star})\n{n}_{0}$, one can rewrite Eq.~(\ref{Eq:expansion_second_term}) as

\begin{align}
    \Delta f^{(1)}_{\rm SSB} \approx &-\frac{\vert \n{v}_{\perp} \vert^{2} }{c D}f^{(0)}_{\rm SSB} +\mathcal{O}(\beta^{3}_{\star}), \label{eq:rewritten_first_term}
\end{align}

\noindent Eq.~(\ref{eq:rewritten_first_term}) is identical to the Shklovskii effect in Eq.~(\ref{Appendix_B:Shklovskii_effect}). 

When calculating the ${\cal F}$-statistic and hence the ${\cal J}$-statistic, the phase of the GW signal as measured in Earth's reference frame, i.e. Eq.~(130) in Ref.~\cite{JaranowskiKrolak1998}, is written in terms of the spin-down parameters $f_{0}^{(k)}$. The latter quantities do not equal  $f^{(k)}_{\rm SSB}$ (defined by Eq.~(\ref{Eq:fSSB_terms_fNS})). Instead they equal the terms in Eq.~(\ref{Eq:phase_SSB_expansion}) proportional to $t^{k}$, for example

\begin{align}
    f_{0}^{(0)} &= f_{\rm SSB}^{(0)}, \\
    f_{0}^{(1)} &= f^{(1)}_{\rm SSB} + \Delta f^{(1)}_{\rm SSB}, \label{Eq:fo^1_in_f^1_SSB_terms}
\end{align}

\noindent and so forth. From Eq.~(\ref{eq:rewritten_first_term}), Eq.~(\ref{Eq:fo^1_in_f^1_SSB_terms}), and $f^{(1)}_{\rm SSB}=\dot{f}_{\star}^{({\rm obs})}$, it is clear that the first spin-down parameter $f_{0}^{(1)}$ entering the calculation of the $\mathcal{F}$-statistic (and hence ${\cal J}$-statistic) is given by

\begin{equation}
    f_{0}^{(1)} = \dot{f}_{\star}^{({\rm obs})}+\dot{f}_{K}.
    \label{eq:f_0^1_spindown_param}
\end{equation}

 \noindent That is, the Shklovskii effect enters \textit{naturally} into the calculation of the ${\cal F}$- and ${\cal J}$-statistics, as laid out in Ref.~\cite{JaranowskiKrolak1998} and implemented in the LALSuite. This exercise can be extended to higher-order frequency derivatives.

\section{Surviving candidates and follow-up}
\label{AppendixC}

\begin{table}[tbp]
\caption{\label{tab:params_Survivors} Terminating frequency bin,  $q^{*}(t_{N_{T}})$, log likelihood, \maxL, detection threshold, $\mathcal{L}_{\rm th}$, and time of ascension, $T_{\rm asc}$, of the surviving candidates. All candidates share the same template for $a_{0}$ and $P$, given by the central values in Table~\ref{tab:Target_Params}.}
\begin{ruledtabular}
\begin{tabular}{cccc}%c
$q^{*}(t_{N_{T}})~(\mathrm{Hz})$ & \maxL & $\mathcal{L}_{\rm th}$ & $T_{\rm asc}~(\mathrm{GPS~time})$ \\ \hline 
$219.6208803$ &  $148.23$ & $148.22$ & 1265652983.7837  \\
248.7994215 &  $150.25$ & $150.24$ & 1265652985.7228  \\
271.3585692 &  $153.38$ & $153.37$ &1265653034.7963  \\
320.2733869 &  $154.31$ & $154.30$ & 1265652922.2337  \\
371.4158598 & $153.82$ & $153.82$ & 1265652985.7228 \\
\end{tabular}
\end{ruledtabular}
\end{table}

\begin{figure*}[ht]
    \includegraphics[scale=0.325, center]{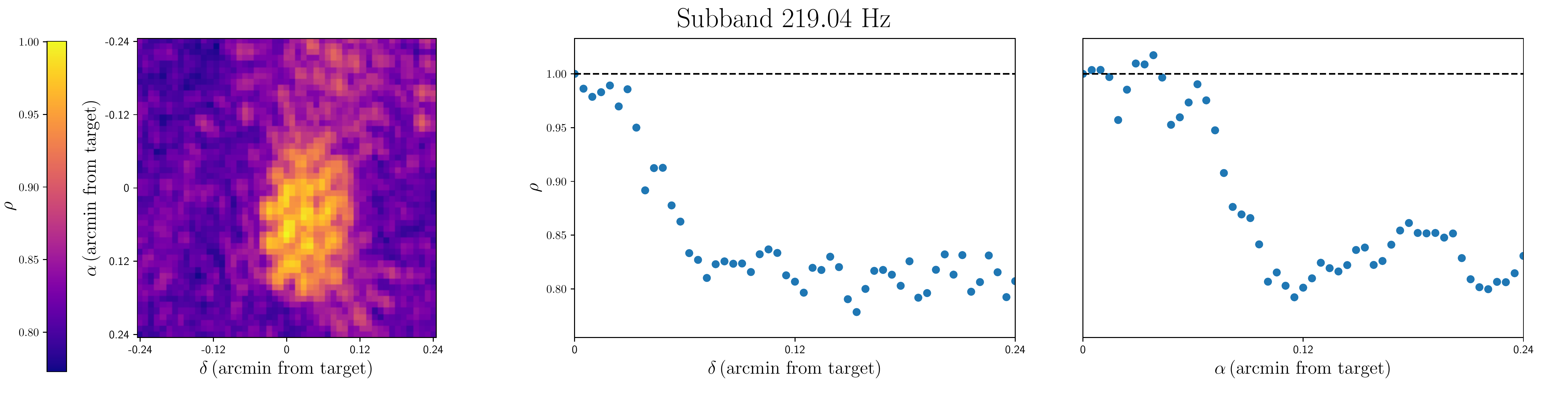}
    \includegraphics[scale=0.325, center]{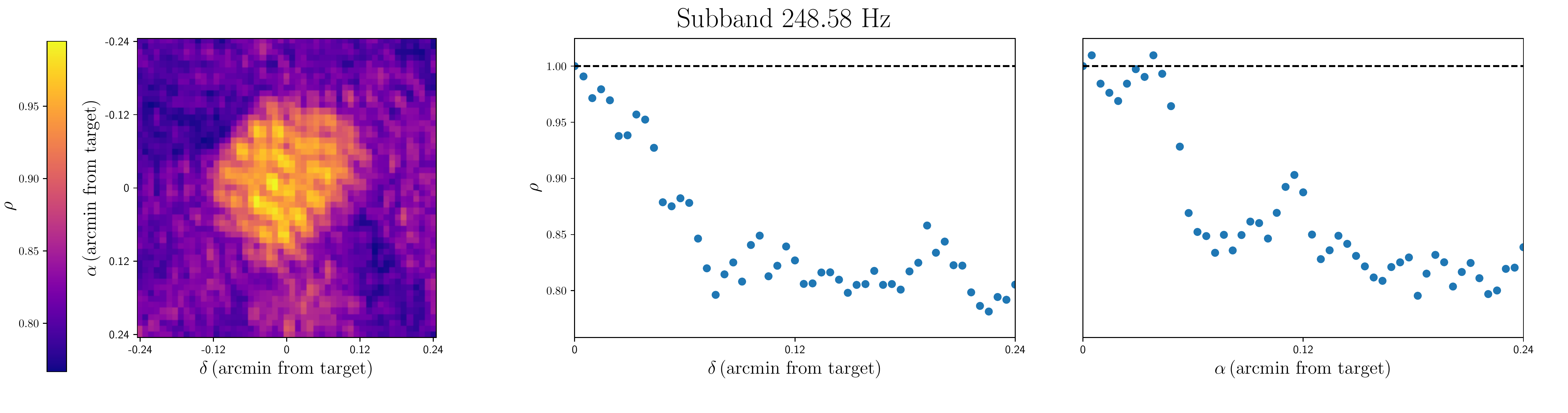}
    \includegraphics[scale=0.325, center]{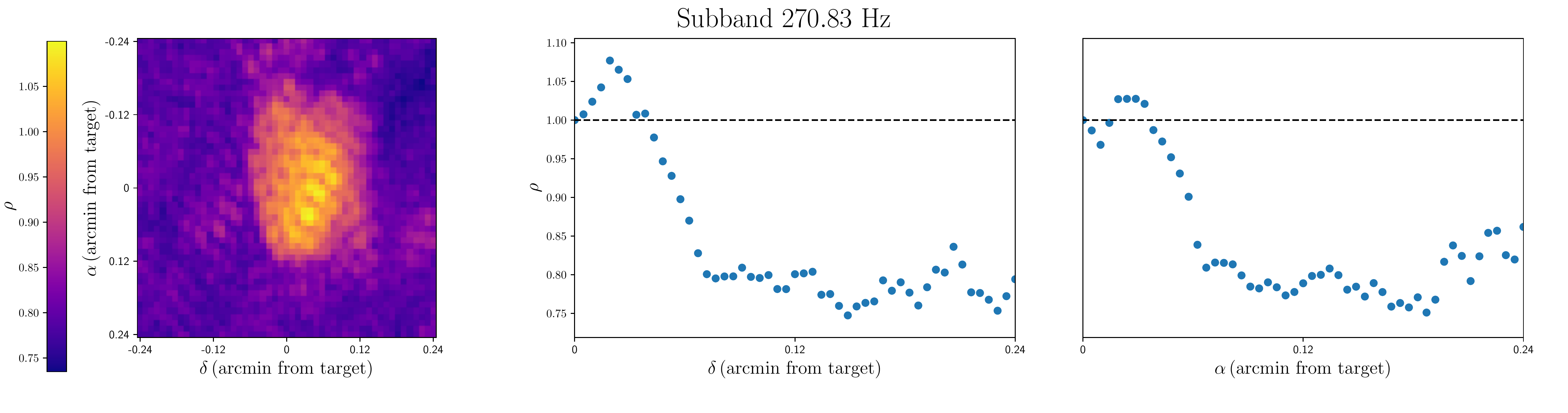}
    \includegraphics[scale=0.325, center]{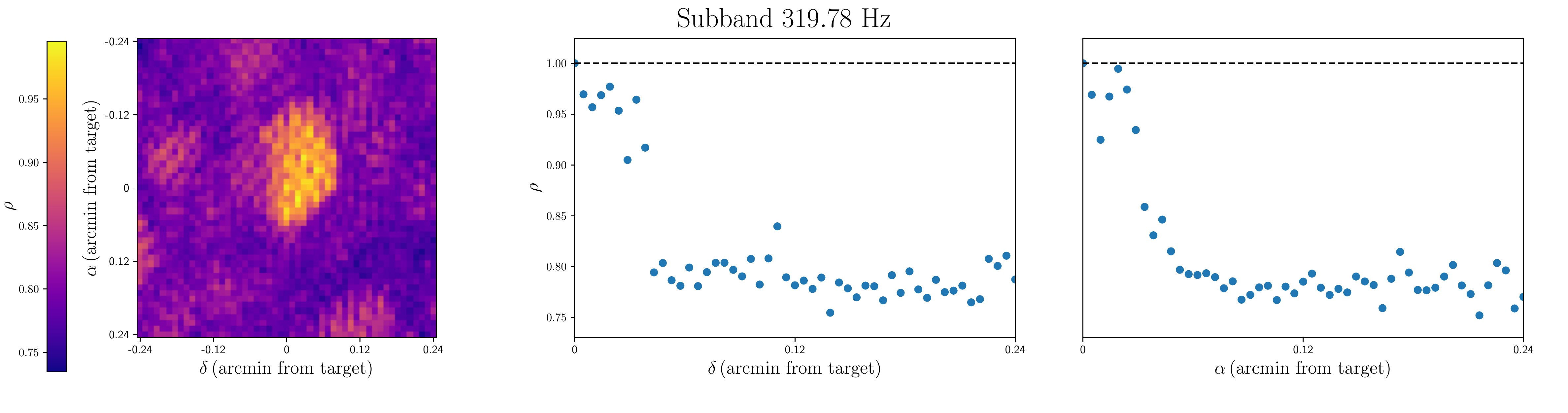}
    \includegraphics[scale=0.325, center]{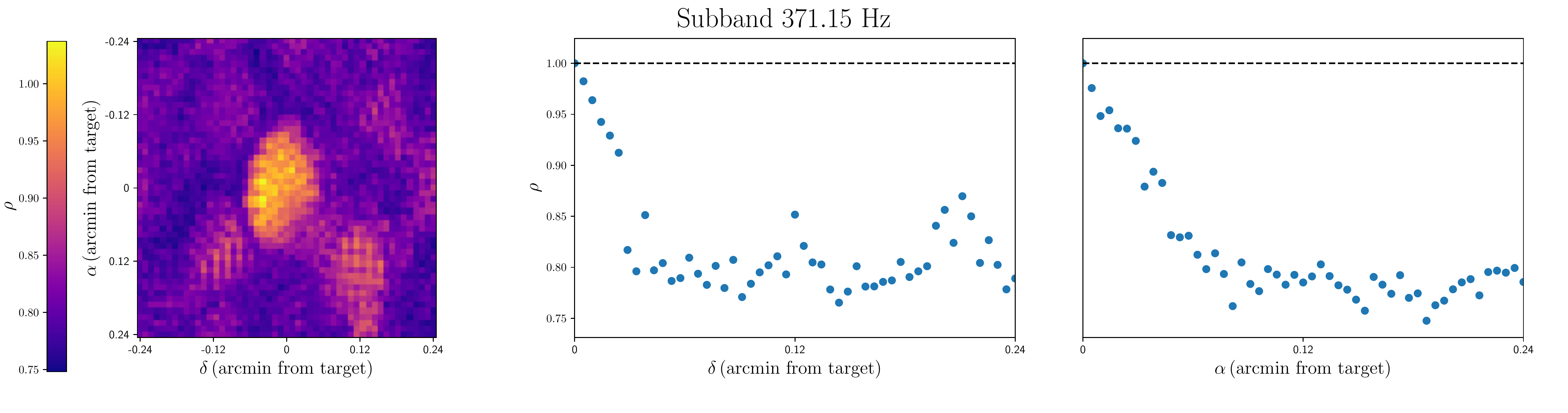}
    \caption{Sky maps of $\rho(\alpha,\delta)={\cal L}(\alpha,\delta)/$\maxL~ (left panel), and slices of $\rho$ at constant $\alpha$ (central panel) and constant $\delta$ (right panel) for the five \Tii~survivors. The sky maps consist of $2601$ regularly spaced sky locations, with each location colored by its $\rho$ value,  spanning a sky area of $0.23\,{\rm arcmin}^{2}$ centered around the source position. For both the central and right panels, $\rho$ is plotted on the vertical axis, while the dashed black line represents the normalized candidate threshold, i.e. $\mathcal{L}_{\rm th}$/\maxL.}   
    \label{fig:dRAdDECSKYMAPS}
\end{figure*}

In this appendix we show how varying the sky position of the surviving candidates, listed in Table~\ref{tab:params_Survivors}, affects $\mathcal{L}$. In general an astrophysical candidate's $\mathcal{L}$ should diminish, as one points away from its original position. Moreover the contours of ${\cal L}(\alpha,\delta)$ should be elliptical, with the orientation of their major axis changing with sky location. Similar tests have been done in previous works, such as in Appendix B.1 of~\cite{TheLIGOScientificCollaborationtheVirgoCollaboration2021} and in Ref.~\cite{JonesSun2022}. Figure~\ref{fig:dRAdDECSKYMAPS} presents the results of the additional follow-up in terms of sky maps and cross-sections of constant-$\delta$ and constant-$\alpha$. 

The sky maps in the left panels of Figure~\ref{fig:dRAdDECSKYMAPS} cover a $0.48\times0.48\,{\rm arcmin}^{2}$ patch of sky centered at the source location. The image uses $51\times51=2601$ evenly spaced grid points. The color of a pixel represents the value of the normalized log likelihood, i.e. $\rho=\mathcal{L}/$\maxL, obtained in that pixel. In general the contours of ${\cal L}(\alpha,\delta)$ for an injection (analogous to the point spread function of an electromagnetic telescope) are elliptical, centered at the source, with orientation and eccentricity depending on $\alpha,\,\delta$, and the orientation of the source \cite{TheLIGOScientificCollaborationtheVirgoCollaboration2021}. None of the ellipses are centered at the source location \footnote{The center of the ellipse is calculated by averaging the position of pixels satisfying $\mathcal{L} > \mathcal{L}_{\rm th}$ (marked by red dots in Figure~\ref{fig:Example_SkyMap_Focii}).}. The closest ones, with centroid offsets $\Delta \alpha=0.006\,{\rm arcsec}~{\rm and}~\Delta \delta=0\,{\rm arcsec}$ and $\Delta \alpha=-0.012\,{\rm arcsec}~{\rm and}~\Delta \delta=0\,{\rm arcsec}$, are associated with the $248.58$ Hz sub-band and $371.15$ Hz sub-band candidates, respectively, which happen to share the same $T_{\rm asc}$. All other ellipses are off-center by more than $1~{\rm arcsec}$. None of the ellipses are centered exactly at the source location even when accounting for \target's proper motion, i.e. $\Delta \alpha =0.12$ arcsec and $\Delta \delta=-0.071$ arcsec.  

Constant-$\delta$ and constant-$\alpha$ cross-sections are graphed in the central and right panels of Figure~\ref{fig:dRAdDECSKYMAPS} respectively. The maximum offset is $0.24\,{\rm arcmin}$ and uses $50$ evenly spaced trials to resolve the rapid change of $\mathcal{L}$. The coordinate offset is graphed on the horizontal axis while $\rho$ is plotted on the vertical axis. The black dashed line, in both panels, represent the normalized candidate threshold $\mathcal{L}_{\rm th}$/\maxL. Only the $371.15$ Hz sub-band candidate shows a monotonic decrease in $\rho$, as both the $\delta$ and $\alpha$ offsets increase, although $\rho$ falls away non-monotonically in every sub-band. The graphs settle around $\rho \approx 0.8$ which is consistent with noise-only $\mathcal{L}$ values. 

\begin{figure*}
    \centering
    \includegraphics[scale=0.7, center]{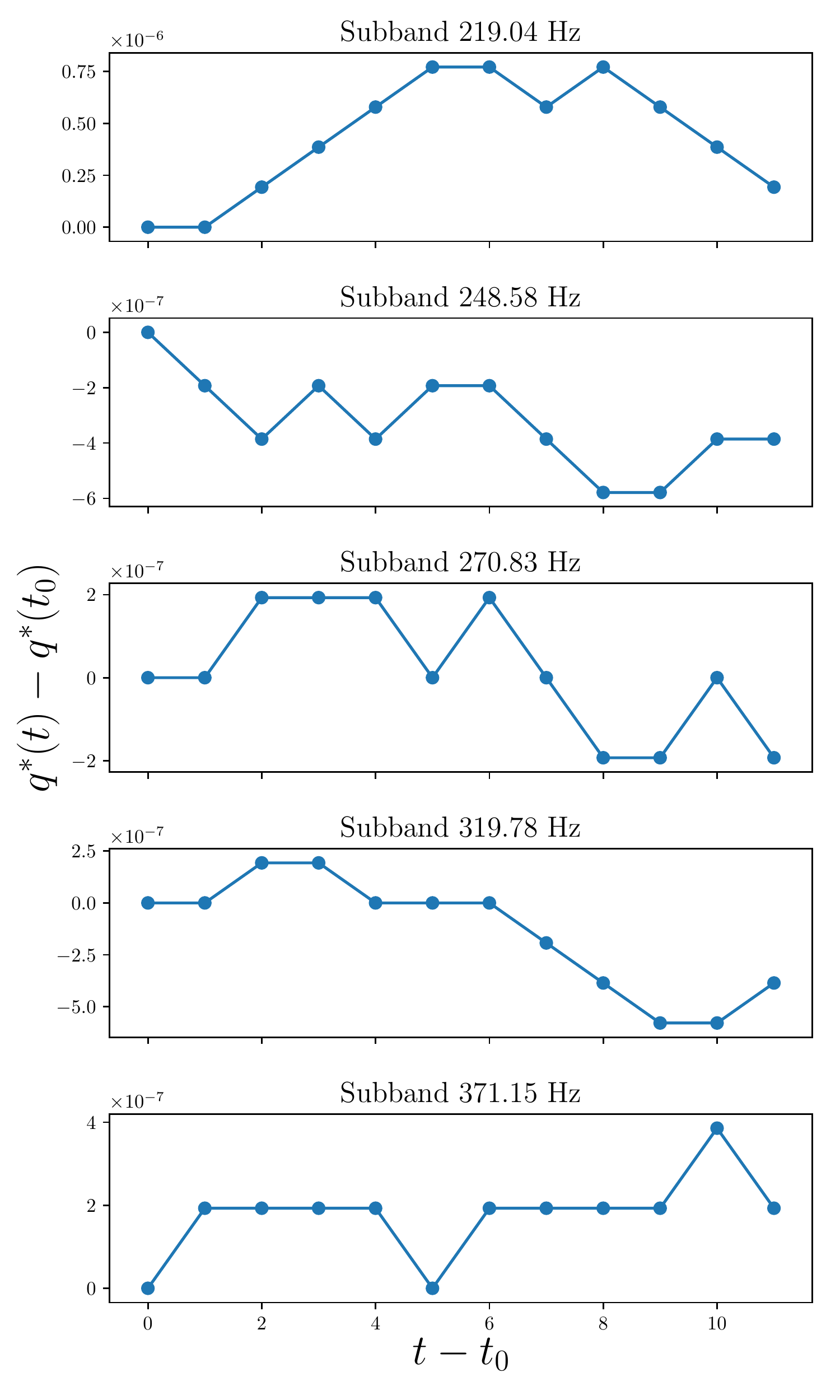}
    \caption{Optimal path for the five \Tii~survivors, measured from the starting frequency bin $q^{*}(t_{0})$, as a function of epoch $t$ (in units of $T_{\rm drift}^{\rm (ii)} = 30 \, {\rm days}$). }
    \label{fig:Paths_Viterbi}
\end{figure*}

For completeness, Figure~\ref{fig:Paths_Viterbi} shows the  optimal path relative to the starting frequency bin, $q^{*}(t_{0})$, for the five \Tii~survivors as a function of the search epoch. There is no clear trend among the optimal paths. In particular the two paths for the candidates sharing the same $T_{\rm asc}$, viz. $248.58$ Hz and $371.15$ Hz, seem to be unrelated.

Given that the number of surviving candidates is consistent with the number of expected false alarms, and that all five are near ${\cal L}_{\rm th}$, we leave further investigations into their origin until newer data sets and/or more sensitive pipelines become available, starting with the fourth LIGO-Virgo-KAGRA observing run.

\newpage
\nocite{*}
\bibliography{only_ads_references,non_ads_references}% Produces the bibliography via BibTeX.

\end{document}